\documentclass[galaxies,article,submit,moreauthors,pdftex, biblatex, 10pt,a4paper]{mdpi}

\firstpage{1} 
\articlenumber{x}
\doinum{10.3390/------}
\pubvolume{xx}
\pubyear{2016}
\copyrightyear{2016}
\externaleditor{Academic Editor: }
\history{Received: date; Accepted: date; Published: date}
 \theoremstyle{mdpi}
 \newcounter{thm}
 \newcounter{ex}
 \newcounter{re}
 \theoremstyle{mdpidefinition}

\Title{EXPLORING THE MAGNETIC FIELD CONFIGURATION IN BL~LAC USING GMVA}

\Author{B. Rani$^{1,2,\dagger}$*, T. P. Krichbaum$^{2}$, J. Hodgson$^{3,}$, S. Koyama$^{2}$, J. A. Zensus$^{2}$, L. Fuhrmann$^{2}$,
A. P. Marscher$^{4}$, S. G. Jorstad$^{4}$ }
\AuthorNames{Firstname Lastname, Firstname Lastname and Firstname Lastname}

\address{%
$^{1}$ \quad NASA Goddard Space Flight Center, Greenbelt, MD 20771, USA \\
$^{2}$ \quad Max-Planck-Institut f{\"u}r Radioastronomie, Auf dem H{\"u}gel 69, 53121, Bonn, Germany\\
$^{3}$ \quad Korea Astronomy and Space Institute, 776 Daedeokdae-ro, Yuseong-gu, Daejeon 34055, Korea \\
$^{4}$ \quad Institute for Astrophysical Research, Boston University, 725 Commonwealth Avenue, Boston, MA 02215, USA}

\corres{Correspondence: bindu.rani@nasa.gov; Tel.: +1-301-286-2531}

\firstnote{NASA Postdoctoral Program (NPP) Fellow} 

\abstract{The high radio frequency polarization imaging of non-thermal emission from active galactic 
nuclei (AGN) is a direct way to probe the magnetic field strength and structure in the immediate 
vicinity of supermassive black holes (SMBHs) and is crucial in testing the jet-launching scenario. 
To explore the the magnetic field configuration at the base of jets in blazars, we took advantage of 
the full polarization capabilities of the Global Millimeter VLBI Array (GMVA). With an angular 
resolution of $\sim$50 micro-arcseconds ($\mu$as) at 86 GHz, one could resolve scales up  to $\sim$450~gravitational radii
(for a 10$^9$ solar mass black hole at a redshift of 0.1). We present here the preliminary results of our study on the blazar BL~Lac.  Our results suggest that on sub-mas scales the core and the central 
jet of BL Lac are significantly polarized with two distinct regions of polarized intensity. We also noted 
a great morphological similarity between the 7mm/3mm VLBI 
images at very similar angular resolution. 
}

\keyword{active galaxies; BL Lacertae object: BL~Lac; jets; GMVA: high-resolution VLBI; magnetic field; polarization}

\begin{document}

\section{High-frequency and high-resolution VLBI}
Polarization study of non-thermal emission from AGN is a direct way to probe the magnetic field strength and
structure in the immediate vicinity  of a black hole, i.e., in a region where plasma is being
injected and accelerated into the main jet stream. The current GMVA observations offer an angular resolution of 50~$\mu$as, which scales down  to $\sim$450 gravitational  radii for 
a 10$^9$~M$_{\odot}$ BH  at a redshift of 0.1. In addition to that, 
Faraday depolarization effects become negligible at millimeter and sub-millimeter radio bands. Therefore, high-frequency 
polarimetric observations are essential in order to have a better understanding of  the role of magnetic field in AGN 
accretion and jet production. 

High-resolution observations have also been proven to be quite important in pinpointing  the radiation processes 
responsible for the $\gamma$-ray emission in blazars \citep{hodgson2016, karamanavis2016, rani2014, rani2015, 
marscher2008}. Magnetic fields appear to play an important role in particle acceleration;  either particles are 
accelerated in high-magnetized environments \citep[via relativistic magnetic reconnection and/or 
magnetoluminescence][]{kagan2015, blandford2015}, or in low magnetized environments via relativistic shocks 
\citep{Marscher1985}.  Understanding magnetic-field configurations is therefore 
essential for probing the high-energy radiation processes.

BL Lacertae, the prototype of the BL Lac class of AGN, is one of the nearest blazars (z=0.068) with the jet
aligned within (6--10)$^{\circ}$ to our line-of-sight, approaching at an apparent flow speed of up to 
$\sim$10c \citep{jorstad2005, cohen2014, cohen2015}. Observing campaigns at 7 mm have shown swings in the direction of the innermost ($\leq$ 1.0 mas)
region of the jet, which has been attributed to either helical instabilities, or jet inlet precession with
proposed periods of 2.3 and 26 years \citep{mutel2005, jorstad2005}. The origin of the observed
position angle swings and the helical motion in the inner jet region is controversial and still enigmatic.
A one-to-one comparison of the jet kinematics with the broad-band flux variability is an important step
towards a solution of the problem. For a recent flare in 2006, a combination of high-resolution images
with the associated broadband flux and optical polarization measurements of the source provided evidence
for a helical magnetic field, well within the jet acceleration zone. There are also growing evidence that
the observed $\gamma$-ray emission is produced in this region, possibly by the interaction of moving with
stationary shock(s) \citep{marscher2008}. Therefore, BL Lac is an excellent candidate to study the relation
between jet formation, $\gamma$-ray emission, shock propagation and polarization variability.

\begin{figure}[t]
\includegraphics[scale=0.38,trim=5 1 5 2.5, clip=true]{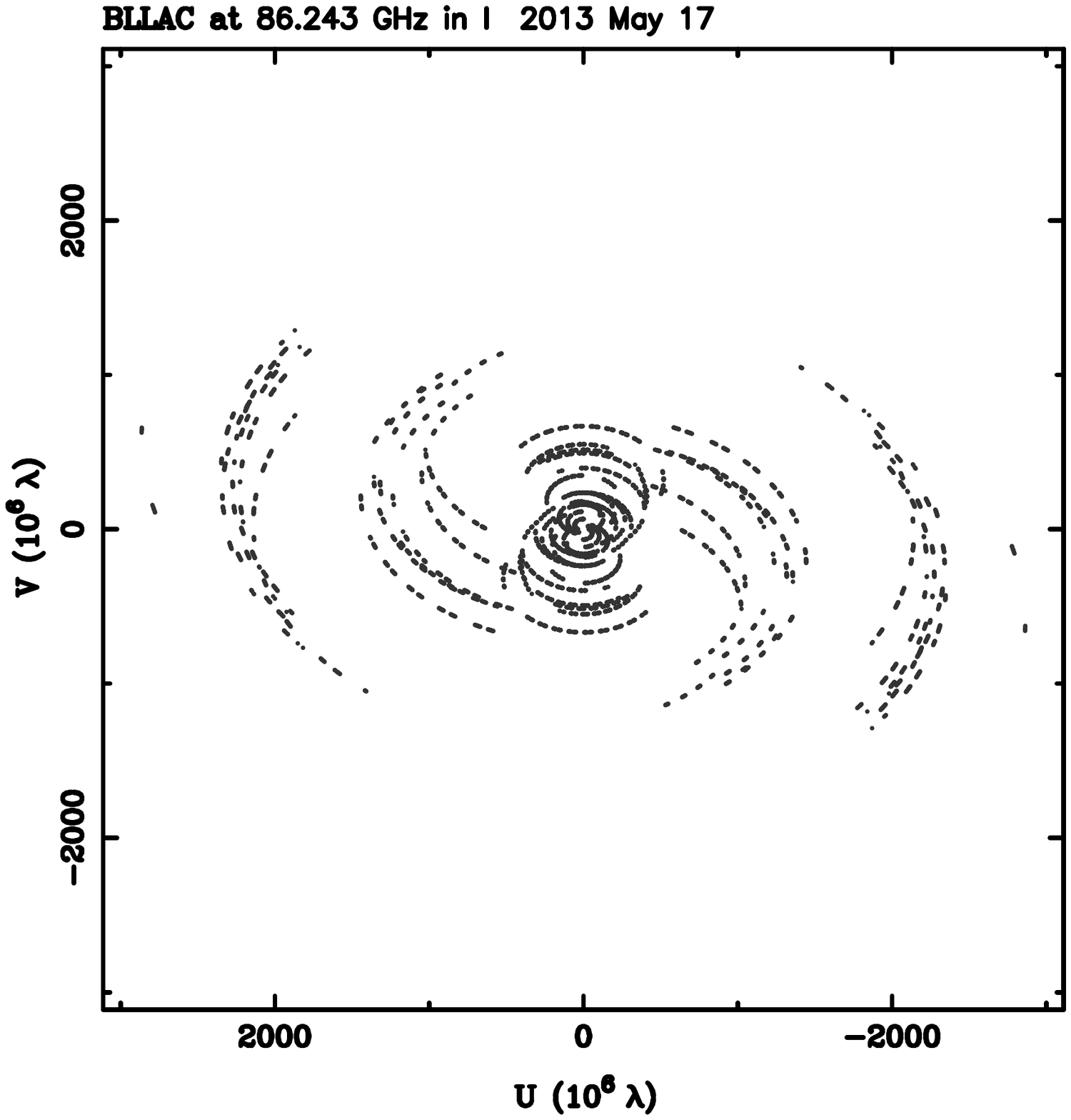}
\includegraphics[scale=0.38,trim=5 1 5 2, clip=true]{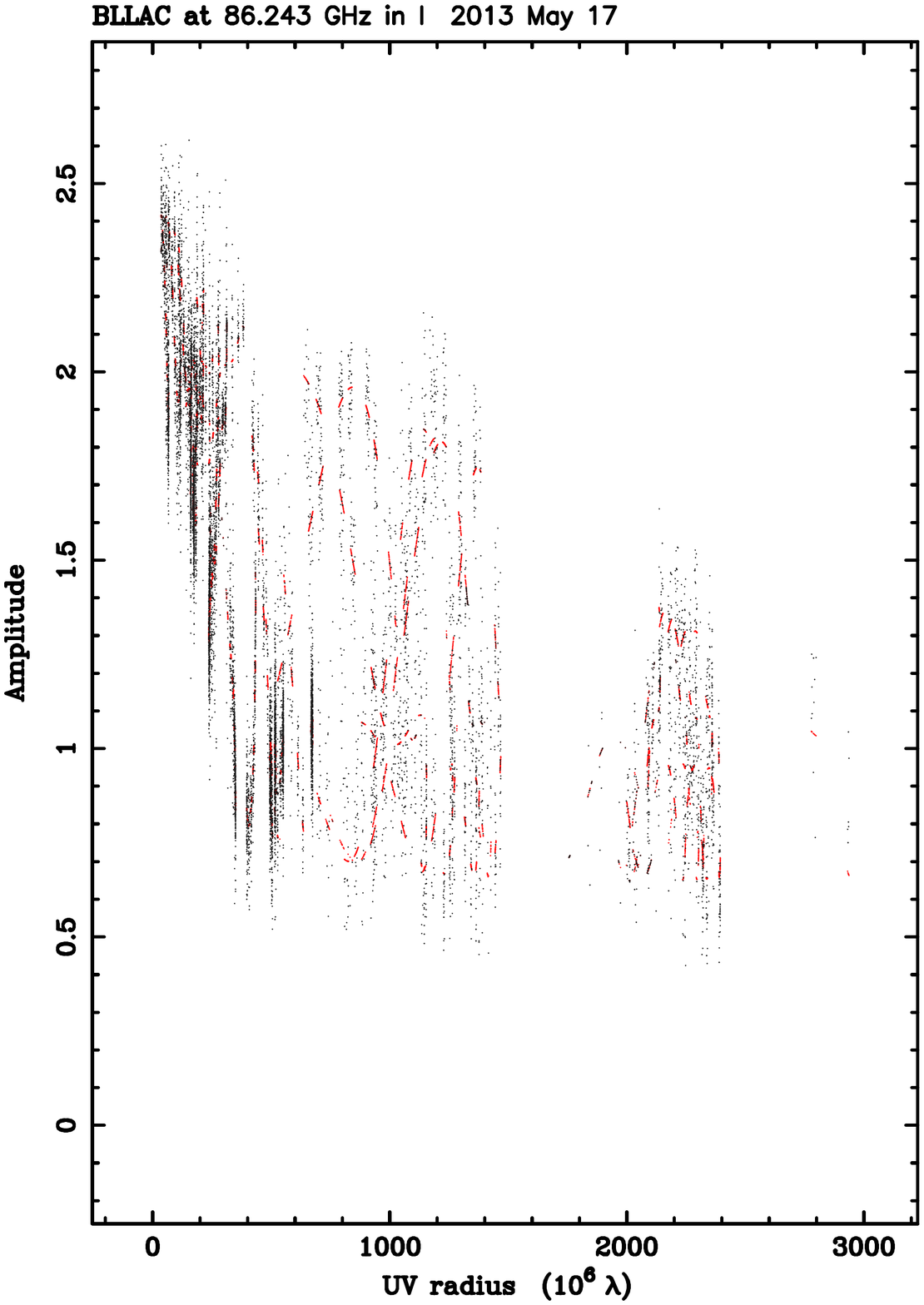}
\caption{Left: UV-plot for BL~Lac at 86~GHz in May 17, 2013. Right: Correlated flux versus projected baseline 
(data in dark-grey and source model in red) .}
\label{fig1}
\end{figure}

\begin{figure}[t]
\includegraphics[scale=0.58,trim=180 0 140 0, clip=true]{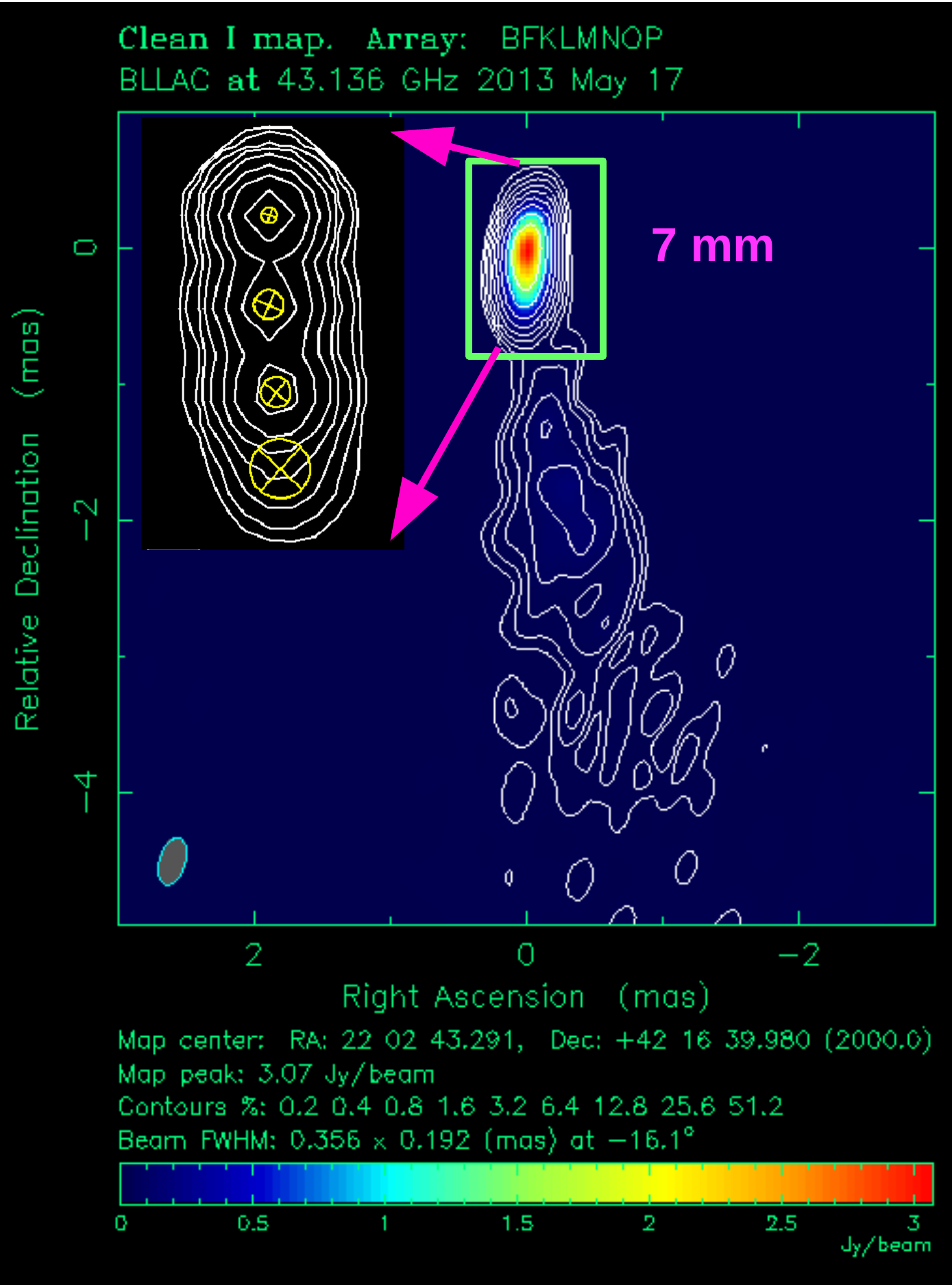}
\includegraphics[scale=0.58, trim=190 0 130 0, clip=true]{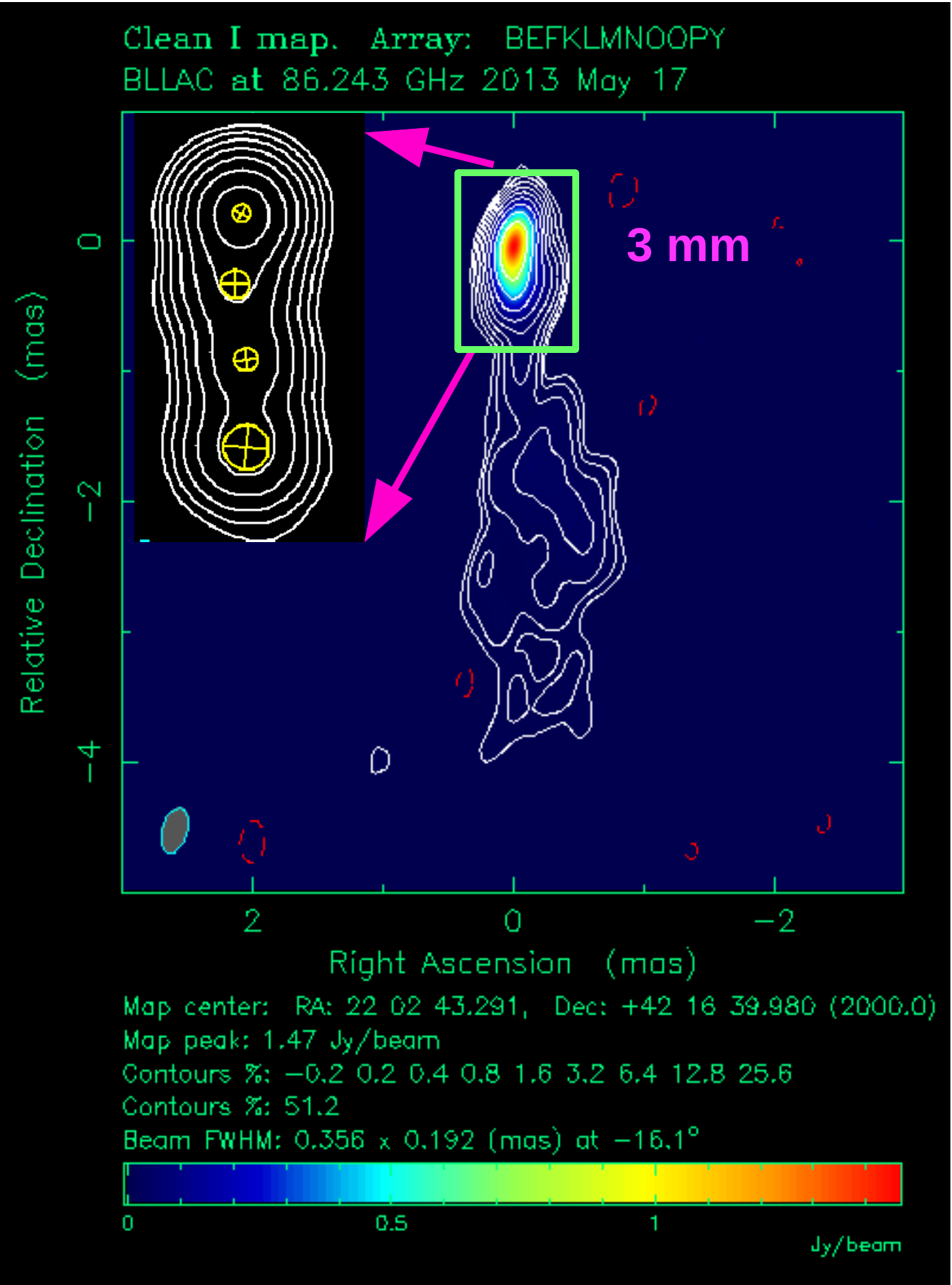}
\caption{Total intensity maps of BL~Lac at 7~mm (left) and 3~mm (right) radio bands. For comparison 
purposes the 3~mm map is convolved with the 7~mm beam. A zoomed version (convolved with a beam size of 0.1~mas) 
of the central region is shown in the upper left corner for each case; the yellow circles are the fitted 
circular  Gaussian components. }
\label{fig2}
\end{figure}

\section{Global 3~mm VLBI observations} 
The VLBI observations of the source were taken during an unprecedented broadband outburst. 
The source was detected at its historic high brightness in cm and mm radio bands   
\citep{Karamanavis2012, wehrle2012}, at
X-ray \citep{Grupe2012} and in the far-infrared bands in December 2012.
On November 2012, the source flux had risen to $\sim$14.6 Jy at 1.3 mm - 
the highest measured since SMA (sub-millimeter array) observations began in 2002. 
On December 5, 2012, BL Lac experienced an X-ray flare several orders of magnitude 
brighter than any previous flare, accompanied 
by bright optical and ultraviolet emission. The source also experienced its brightest 
$\gamma$-ray flare measured by the {\it Fermi}-LAT (Large Area Telescope)
\citep[Photon Flux at E$>$100 MeV $\sim$ 2$\times$10$^{-6}$ ph cm$^{-2}$ s$^{-1}$,][]{cutini2012}. 
We took advantage of this unique opportunity to search for
related structural variations in the core region and on scales of a few ten micro-arcseconds. 

To achieve the necessary high resolution and image fidelity, 
we conducted immediate VLBI observations in dual polarization mode at 22, 43, and 86~GHz radio frequencies in 2013. 
The bright blazars S5 0716+714, 1749+096, 3C 454.3, and 3C 345 were used as calibrators. 
In addition to the VLBA, we had the participation of Pico Veleta and Effelsberg (both LCP/RCP) and Yebes (LCP). 
As a result, we had a factor of two increase in angular resolution and a factor of three gain in 
the sensitivity. Data calibration and imaging were performed using the standard tasks in AIPS and difmap. 
Figure \ref{fig1} (left) shows the UV-coverage of BL~Lac for the 3~mm data taken in May 2013. The correlated flux 
versus baseline length plot for the same is shown in the right panel.

Determining the D-terms (often refereed to as polarization leakage) is 
one of the most challenging aspects of high-frequency polarization imaging. So far there is no optimal 
way to get an accurate estimation of them. We use the task LPCAL to estimate the D-terms of each antenna for 
a given experiment. For a consistency check, we compare the estimated D-terms for different sources in a given 
experiment and also for the same source in different experiments. Except for a few stations, we noticed 
that the variations in D-term values were $\leq$15$\%$. For stations like MK, the D-term variations were 
of the order of 20$\%$. The last step is to correct the EVPA (electric vector polarization angle)  of the 
source. To do so, we compare the EVPA of the map convoled with a larger beam (especially of the optically thin region) 
with the close-in-time  single dish EVPA measurements. The difference between the two is then applied to the final image. 
A detailed overview of mm-VLBI polarization calibration and imaging can be found in \cite{marti2012}.  
In the following section, we present the preliminary results of our 
study.

\begin{figure}
\center
\includegraphics[scale=1.0,trim=90 150 30 85, clip=true]{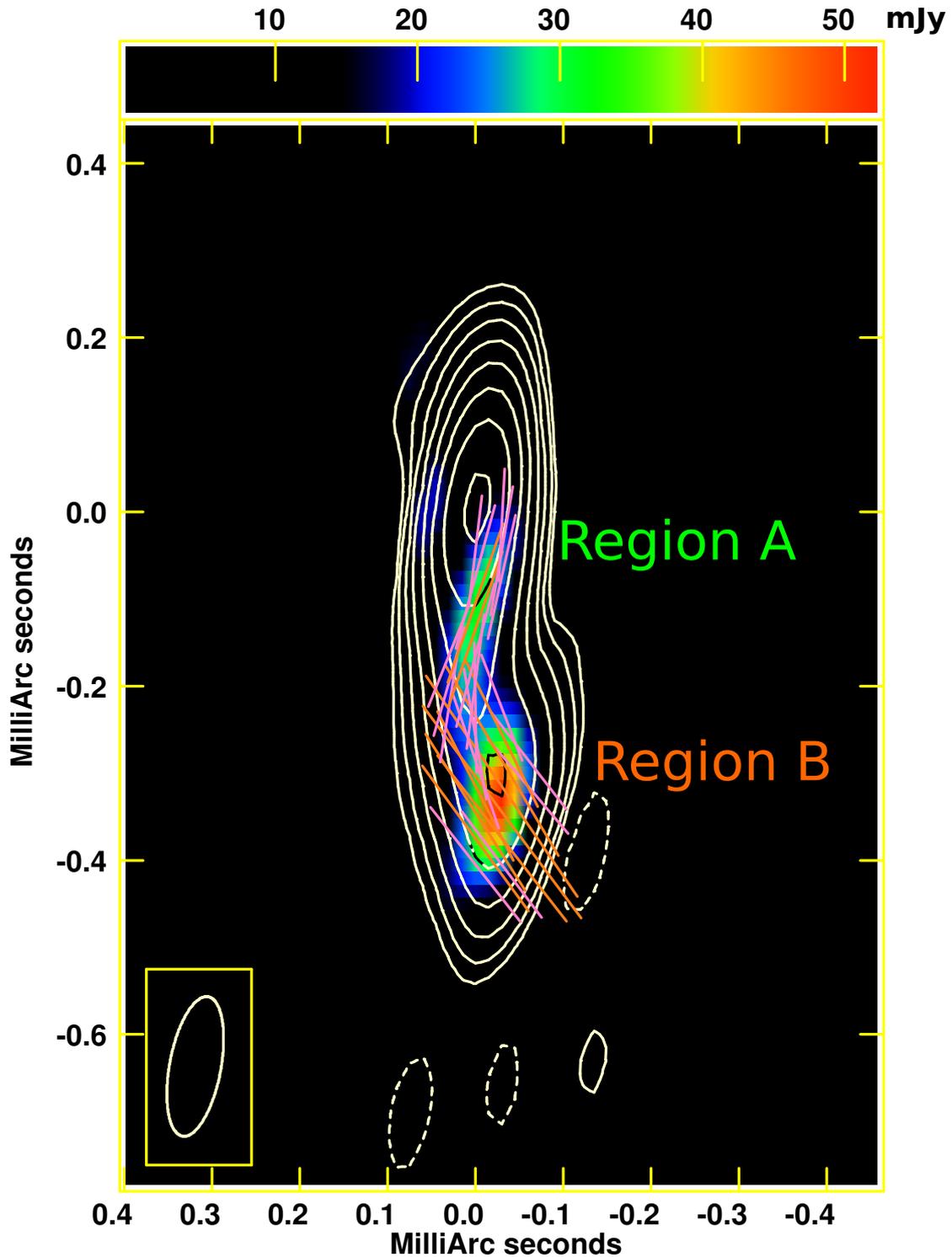}
\caption{High-resolution intensity and polarization image of BL~Lac at 86~GHz (3~mm) obtained 
in May 17, 2013 (beam FWHM: 0.172$\times$0.0748 (mas) at $-$8.38$^{\circ}$). Contours 
represent total intensity and the peak flux is 1.04~Jy/beam; contour levs 
are (-0.07, 0.07, 0.14, 0.28, 0.56, 1.12, 2.24, 4.48, 8.96, 17.92, 35.84, 71.68)$\%$ of the peak flux. The 
polarized intensity is shown via the color scale. The solid ticks mark the EVPA direction.  }
\label{fig3}
\end{figure}

\section{Results}
Figure \ref{fig2} shows the 7~mm (left) and 3~mm (right) total intensity maps of BL~Lac for an experiment 
conducted in May 17, 2013. The 3~mm map is convolved with the 7~mm beam to compare the extended jet morphology. 
 The two images look very similar. However the extended jet region is much fainter at 3mm than at 7mm.  In the top left corner, we show a zoomed version of the inner 
jet region ($<$0.4~mas), which corresponds to a linear distance of $<$5,000 gravitational radii in the source frame.  
The inner jet region can be well described by four circular Gaussian components 
(yellow circles) both at 43 (7~mm) and 86~GHz (3~mm) frequencies. This implies a  similarity  
of jet morphology in the inner region as well.   The  simultaneous multi-frequency observations will be used 
to determine the spectral turnover for individual jet knots. Given their spectral indices and VLBI sizes, 
the magnetic field at different jet locations can be determined. The estimated magnetic field strengths at 
different separations from the 
jet apex will then be used to constrain the magnetic field strength at the jet apex. Recent evidence suggests that 
the central black holes in jetted AGN are surrounded by magnetically arrested disks \citep{tchekhovskoy2011}, which 
implies high magnetic fluxes close to the central engines. Our observations will provide observational constraints 
on the magnetic field strength and configuration close to the super-massive black holes.

In Fig.\ \ref{fig3} we show as an example the 86~GHz polarization map of BL~Lac convolved with its natural beam. 
The color scale represents the polarized intensity which is plotted over the total intensity contours (I-map), 
and the solid lines mark the EVPA  direction.    Two regions of polarized intensity are of 
particular interest here; we mark these regions as Region~A and Region~B. Region~A is a bright polarized region 
elongated from the core up to a distance of $\sim$0.2~mas. The EVPA in this region is roughly inclined by 
40-50$^{\circ}$ to the jet axis. In region B we noticed an almost 90$^{\circ}$ change in the EVPA direction 
in comparison to region A.  Region B roughly extends from a distance of 0.2 to 0.5~mas and it is even more bright 
than region~A. The polarized intensity scale roughly follows the total intensity contours in the two regions. 
 Region A and B have a percentage polarization of 14$\%$ and 30$\%$, respectively.
The change in the EVPA direction could either be due to the presence of helical magnetic 
field, which is a natural consequence of rotation of the central accretion disk and outflow, or it could simply be 
due to the presence of multiple oblique shocks.   A comparison of multi-epoch polarization maps will allow us to 
resolve this question.

\section{Summary and outlook}
Understanding the physical processes happening close to the central engines and their connection to 
the jet activity and to the broadband flaring activity in blazars are the key challenges in active 
galactic nuclei physics. Ultra-high angular resolution mm-VLBI observations are the most feasible ways 
to answer these questions. 
We demonstrated the high-frequency and high-resolution polarization imaging capabilities of 
the GMVA. In the very near future, ALMA will participate in the observations at 3~mm and 1.3~mm. 
This will significantly enhance the imaging and polarization capabilities of global mm-VLBI observations. 


\vspace{6pt} 


\acknowledgments{This research was supported by an appointment to the NASA Postdoctoral Program
at the Goddard Space Flight Center, administered by Universities Space Research Association
through a contract with NASA. BR acknowledges the help of Dave Thompson for his comments on the 
manuscript. 
This research has made use of data obtained with the Global Millimeter VLBI Array
(GMVA), which consists of telescopes operated by the MPIfR, IRAM, Onsala, Metsahovi, Yebes
and the VLBA. The data were correlated at the correlator of the MPIfR in Bonn, Germany.
The VLBA is an instrument of the National Radio Astronomy Observatory, a facility of the
National Science Foundation operated under cooperative agreement by Associated
Universities, Inc.}

\authorcontributions{BR, TK, AM, SJ, AZ, and JH proposed and conducted the observations. BR and SK analyzed 
the data. The manuscript is written by BR. }

\conflictofinterests{The authors declare no conflict of interest.} 


\end{document}